\documentclass[
preprint
,showpacs
,showkeys
,nofootinbib
,aps
,amsfonts
,amssymb
,pdftex 
]{revtex4-1}

\usepackage{graphicx} 
\usepackage{amsmath, amssymb, bm} 
\usepackage[dvips]{color}

\usepackage{psfrag,epsfig}
\usepackage{natbib}
\usepackage{hyperref}

\def\){\right)} 
\def\({\left(} 
\def\]{\right]} 
\def\[{\left[}

\def\nC{$^{14}\mathrm{C}(n, \gamma)^{15}\mathrm{C}$}

\def\P { \left(\frac{\stackrel{\rightarrow}{\nabla}}{M_c}-\frac{\stackrel{\leftarrow}\nabla}{M_n}\right)}

\def\erasepar#1{}

\begin{document}

\title{
Radiative Neutron Capture on Carbon-14 in Effective Field Theory}

\author{%
Gautam Rupak}
\email{grupak@u.washington.edu}

\author{%
Lakma Fernando}
\email{nkf22@msstate.edu}

\author{%
Akshay Vaghani}
\email{av298@msstate.edu}

\affiliation{Department of Physics $\&$ Astronomy and 
HPC$^{2}$ Center for Computational Sciences, 
Mississippi State
University, Mississippi State, MS 39762, U.S.A. }

\begin{abstract}
The cross section for radiative capture   of neutron on carbon-14 is calculated using the model-independent formalism of halo effective field theory. The dominant contribution from E1 transition is considered, and the cross section is expressed in terms of elastic scattering parameters of the effective range expansion.   Contributions from both resonant and non-resonant interaction are calculated.  Significant interference between these leads to a capture contribution that deviates from simple Breit-Wigner resonance form.

\end{abstract}


\maketitle

\section{Introduction}\label{sec_intro}

The radiative capture of neutron on carbon-14  \nC~ plays an important role in astrophysics.  It is part of the neutron induced carbon-nitrogen-oxygen (CNO) cycle in the helium burning layer of asymptotic giant branch stars and  in the core helium burning of massive stars~\cite{Wiescher1999}. These neutron induced reactions can lead to appreciable changes in the  CNO abundances.  \nC~ is the slowest reaction in the cycle and leads to substantial enrichment of $^{14}$C abundance~\cite{Wiescher1999}. In astrophysical scenarios involving inhomogeneous Big Bang Nucleosynthesis,  the slow \nC~ reaction acts as a bottle neck  in   the production of heavier nuclei $A>14$~\cite{Wiescher1990,Kajino1990}.  The \nC~ cross section has been measured in direct capture experiments~\cite{Beer1992,Reifarth2005,Reifarth2008} , and also extracted indirectly from Coulomb dissociation  data~\cite{Horvath2002,Pramanik2003, Nakamura2003,Nakamura2009}. Interpretation of  Coulomb dissociation data for the capture rate requires careful treatment of the parent $^{15}$C and daughter $^{14}$C nuclei in the strong Coulomb field of a heavy nucleus besides the nuclear interactions~\cite{Baur:1986pd, PhysRevC.78.011601,PhysRevC.80.024608}.   
Coulomb dissociation provides an alternate method to estimate the direct capture reactions involving radioactive isotopes that are often difficult to measure experimentally.  The \nC~ provides an opportunity to compare and contrast the capture rates that are obtained from direct capture measurements and Coulomb dissociation data~\cite{PhysRevLett.96.162501,PhysRevC.78.011601}. Developing theoretical methods for radiative capture reactions is important for ongoing experimental efforts, and those planned at FRIB~\cite{FRIB}.  

We calculate the radiative capture  \nC~ cross section at low-energies using halo effective field theory (EFT)~\cite{Bertulani:2002sz,Bedaque:2003wa}.  This reaction has been calculated before in other theoretical formulations such as 
Refs.~\cite{Wiescher1990,PhysRevC.78.011601,Huang:2008ye,Descouvemont2000}. 
Halo EFT has been used to study $s$-wave alpha-alpha resonance~\cite{Higa:2008dn,*Gelman:2009be} and three-body halo nuclei~\cite{Canham:2008jd,*Canham:2009xg}. 
Recently it has been used to calculate electromagnetic transitions and transition probability strength in one-neutron halo $^{11}$Be~\cite{Hammer:2011ye,*Phillips:2010dt}, radiative neutron capture on $^7$Li~\cite{Fernando:2011ts,Rupak:2011nk}, 
and proton-$^7$Li interaction in coupled-channel extension~\cite{Lensky:2011he}.   In EFT,  the cross section is  expressed as an expansion in the small ratio of low-energy physics scale $Q$ of interest over the high-energy physics scale $\Lambda$ that involves short distance physics not relevant at low-energy.  EFT provides a model-independent framework for calculations whose accuracy can be systematically improved as long as there is a clear separation between the energy scales, $Q\ll \Lambda$.  We consider center-of-mass (c.m.) energies $\lesssim 2$ MeV,  corresponding to momenta $p\lesssim 60$ MeV, that is below the threshold for the excited states of  $^{14}$C nucleus (or neutron).  As such in the EFT, the neutron and $^{14}$C core are treated as inert point-like particles.  The ground state of $^{15}$C, identified as  $J^\pi=\frac{1}{2}^+$, has a neutron separation energy $B$ of only $1.218$ MeV that correspond to a binding momenta of $\gamma =\sqrt{2\mu B}\approx 46.21$ MeV, where $\mu$ is the neutron-$^{14}$C reduced mass.  In nuclear structure calculations 
the ground state of 
$^{15}$C can be considered a single neutron halo bound to a $^{14}$C core.  Then in the single-particle approximation, it is described as a $^2S_{1/2}$ state of $n+{}^{14}$C.  We use the spectroscopic notation $^{2S+1}L_J$ with $S$ the spin, $L$ the orbital angular momentum and $J$ the total angular momentum.  The momenta $p$, $\gamma$ are the soft scale $Q$.  The energy threshold for the excited states of  $^{14}$C, pion physics, etc., is identified with the hard scale $\Lambda\sim 100-200$ MeV. 

At low-energy, the capture from lower partial wave initial states should dominate. However,  neutron capture from initial $s$-wave state to the ground state through M1 transition is suppressed (at one-body current level) due to the orthogonality of the continuum and bound state wave functions.   The lowest multipole transition to the ground state is through E1 transition from the initial $p$-wave states $^2P_{1/2}$ and $^2P_{3/2}$.  We note that transition from the initial $s$- and $p$-wave states to the excited state of $^{15}$C $J^\pi=\frac{5}{2}^+$ is possible.  However, transitions to the excited state has been found to be a small contribution to the total capture rate~\cite{Wiescher1990,Descouvemont2000,PhysRevC.80.034611}.  We ignore such contributions in this calculation where we concentrate on the dominant effects.  

The paper is organized as follows. In section ~\ref{sec_theory} we introduce the basic theory and the interactions necessary for the \nC~ cross section calculation.  The Lagrangian for the $s$- and $p$-wave interaction of neutron and carbon-14 is presented. We describe how the EFT couplings can be constrained from data.  The E1 capture cross section is calculated in section ~\ref{sec_results}.  We consider both direct capture and Coulomb dissociation data.  EFT couplings are constrained to reproduce the available data. From the analysis,  we formulate a power counting for estimating the sizes of the couplings and the various EFT contributions.     In section ~\ref{sec_conclusions} we present our conclusions.  

\section{Formalism}\label{sec_theory}
\begin{figure}[thb]
\begin{center}
\includegraphics[width=0.47\textwidth,clip=true]{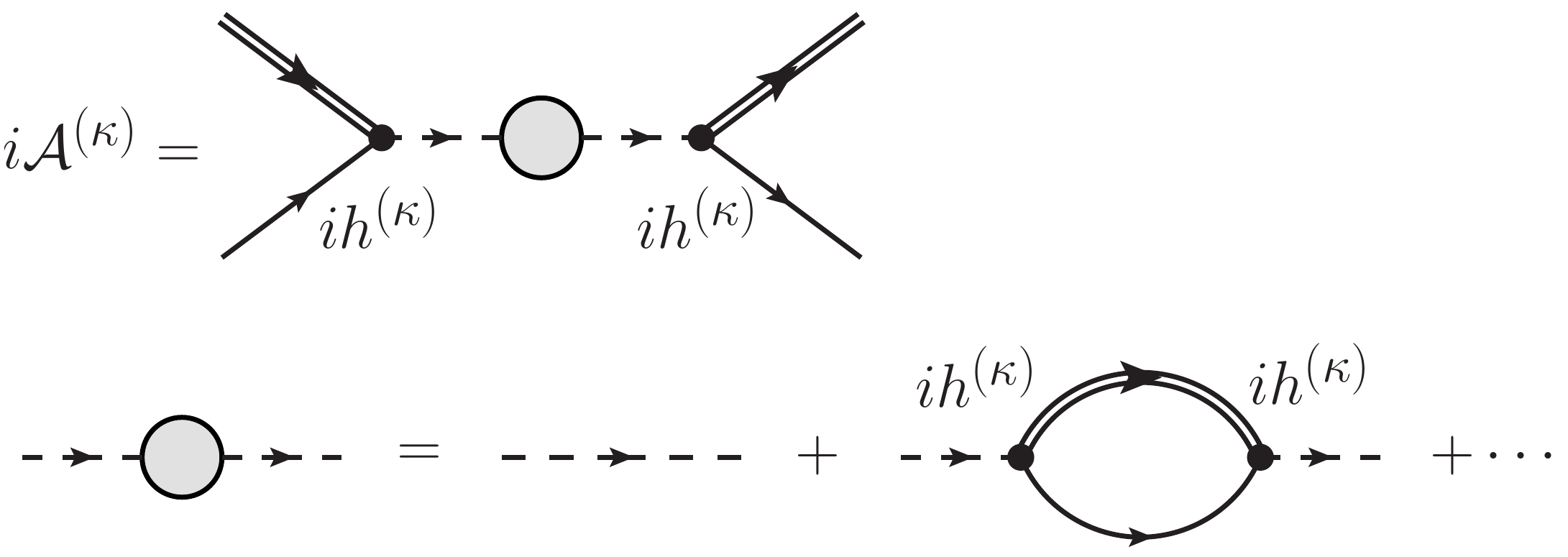} 
\end{center}
\caption{\protect Elastic scattering amplitudes $\mathcal A^{(\kappa)}$ in $s$- and $p$-waves. 
Double line is the $^{14}$C propagator, single line the neutron propagator, 
dashed line the bare dimer propagator. $\kappa=0,1,2$ corresponds to $^2S_{1/2}$, $^2P_{1/2}$ and $^2P_{3/2}$ channels, respectively. 
}
\label{fig:scattering}
\end{figure}

The construction of the EFT for \nC~ require description of the $n+^{14}$C bound state in the $^2S_{1/2}$ channel, and the initial state interaction of $n+^{14}$C in the $^2P_{1/2}$ and $^2P_{3/2}$ channels.   The interaction in the $^2S_{1/2}$ channel is written as
\begin{align}\label{eq:Ls}
\mathcal L_s = \phi_\alpha ^\dagger\[\Delta^{(0)} +i\partial_0+\frac{\nabla^2}{2M}\]
\phi_\alpha+h^{(0)} \[\phi_\alpha^\dagger (N_\alpha C)+\operatorname{h.c.}\], 
\end{align}
where $\phi_\alpha$ is an auxiliary field with a spin index $\alpha$, $N_\alpha$ is the neutron field and $C$ is the carbon-14 scalar field.  $M=M_n+M_c$ with neutron mass $M_n=939.6$ MeV and $^{14}$C core mass $M_c=13044$ MeV. Using the equation of motion for the $\phi$ field, it can be integrated out of the theory in Eq.~(\ref{eq:Ls}), and the interaction Lagrangian written entirely in terms of  four-particle neutron carbon-14 interactions.  The non-relativistic $s$-wave amplitude is calculated from the diagrams in 
Fig.~\ref{fig:scattering}.  We get
\begin{align}\label{eq:s-wave}
i\mathcal A_0(p)=-i h_0^2 D_\phi(\frac{p^2}{2\mu},0), = -\frac{i [h^{(0)}]^2}{\Delta^{(0)}+p^2/(2\mu)+\mu [h^{(0)}]^2(\lambda+i p)/(2\pi)},
\end{align}
where the dressed $\phi$ propagator is
\begin{align}
i D_\phi(p_0,\bm{p})=&\frac{i}{\Delta^{(0)}+p_0-p^2/(2M)+i [h^{(0)}]^2 f_0(p_0,\bm{p})},\\
f_0(p_0,\bm{p})= &-i 2\mu\(\frac{\lambda}{2}\)^{4-D}\int 
\frac{d^{D-1}\bm{q}}{(2\pi)^{D-1}}\frac{1}{q^2- 2\mu p_0 +\mu p^2/M -i 0^+}\nonumber\\
=& 
-\frac{i\mu}{2\pi}(\lambda-\sqrt{-2\mu p_0 +\mu p^2/M-i 0^+}), \nonumber 
\end{align}
with $\lambda\sim Q$ the renormalization scale.
We use the power divergence subtraction scheme where divergences in space-time dimensions $D=4$ and lower are subtracted~\cite{Kaplan:1998tg}. 
In Eq.~(\ref{eq:s-wave}), we iterate the interaction to all order to describe a $s$-wave bound state.   At low energy matching the EFT amplitude Eq.~(\ref{eq:s-wave}) to the effective range expansion (ERE)
\begin{align}
i\mathcal A_0(p)=\frac{2\pi}{\mu}\frac{i}{p\cot\delta_0-i p}\approx \frac{2\pi}{\mu}\frac{i}{-\gamma+\rho(p^2+\gamma^2)/2-i p},
\end{align}
we get
\begin{align}
\frac{2\pi\Delta^{(0)}}{\mu [h^{(0)}]^2}+\lambda=&\gamma-\frac{1}{2}\rho\gamma^2,\\
-\frac{2\pi}{[h^{(0)}]^2\mu^2}=&\rho , \nonumber
\end{align}
where $\mu= M_n M_c/(M_n+M_c)$ is the reduced mass, $\gamma\approx 46.21$ MeV  is  the $^{15}$C ground state binding momentum and $\rho$ is the effective range in $s$-wave. There is no experimental constraint on the value of $\rho$. \emph{A priori} it is not clear if the effective range $\rho$, which has the dimension of length, should scale with the short distance (high-energy)  scale $\rho\sim 1/\Lambda$  or with the long distance (low-energy) scale $\rho\sim1/Q$. 
If its the former, $\rho$ is a next-to-leading order  (NLO) correction whereas if its the latter, its a leading order (LO) contribution in EFT.  

To describe the incoming $^2P_{1/2}$ and $^2P_{3/2}$ states we consider a Galilean invariant form consisting of the relative neutron and $^{14}$C core velocity $\bm{v}_C-\bm{v}_N$, and the neutron field $N_\alpha$ and the scalar carbon-14 field $C$ . In particular we want to project a generic tensor $\psi_i^\alpha$ with a vector index $i=1,2,3$ for the $p$-wave and a spin index $\alpha=1,2$ for the neutron spin into the total angular momentum  $J=1/2$ piece and  $J=3/2$ piece.  This can be done as 
\begin{align}
\psi_i^\alpha =\frac{1}{3}(\sigma_i\sigma_j)^{\alpha\beta}\psi_j^\beta +
\[ \delta_{i j}\delta^{\alpha\beta}-\frac{1}{3}(\sigma_i\sigma_j)^{\alpha\beta}\]\psi_j^\beta,
\end{align}
where the two pieces are the irreducible forms representing   the $^2P_{1/2}$ and $^2P_{3/2}$ states respectively.  $\sigma_i$ are the Pauli matrices. 
Thus the $p$-wave interaction in the EFT can be written as
\begin{align}\label{eq:Lp}
\mathcal L_p={\chi_i^{\alpha, \eta}}^\dagger\[\Delta^{(\eta)}+i\partial_0 +\frac{\nabla^2}{2 M}\] \chi_i^{\alpha,\eta} 
+\sqrt{3} h^{(\eta)}
[{\chi_i^{\alpha,\eta}}^\dagger P_{i k}^{\alpha\gamma,\eta} N_\gamma\P_k C +\operatorname{h.c}],
\end{align}
where $\eta=1,2$ corresponds to the $^2P_{1/2}$ and $^2P_{3/2}$ channels respectively.  These particular $p$-wave channels in $^{11}$Be were also studied in Refs.~\cite{Hammer:2011ye,*Phillips:2010dt}.  The auxiliary field $\chi_i^\alpha$ plays in $p$-wave a  role  similar to the $\phi_\alpha$ field in $s$-wave earlier in Eq.~(\ref{eq:Ls}).   The projectors  $P_{i j}^{\alpha\beta,\eta}$ in Eq.~(\ref{eq:Lp})  are 
\begin{align}\label{eq:PP}
P_{ij}^{\alpha\beta,1} = &\frac{1}{3}(\sigma_i\sigma_j)^{\alpha\beta}, \\
P_{ij}^{\alpha\beta,2} =&\delta_{i j}\delta^{\alpha\beta}-\frac{1}{3}(\sigma_i\sigma_j)^{\alpha\beta}. \nonumber
\end{align}
The $p$-wave elastic scattering amplitude is given by a set of diagrams similar to the $s$-wave amplitude, Fig.~\ref{fig:scattering}.  We get
\begin{align}\label{eq:p-wave}
i\mathcal A_1^\eta(p)= - [h^{(\eta)}]^2 \frac{k^2}{\mu^2} i D_\chi^\eta(p^2/(2\mu),0)
=\frac{2\pi}{\mu} 
\frac{i p^2}{-\frac{2\pi\mu\Delta^{(\eta)}}{[h^{(\eta)}]^2}
-\frac{\pi\lambda^3}{2} -\left(\frac{3 \lambda}{2} 
+\frac{\pi}{[h^{(\eta)}]^2 }\right)p^2 -i p^3},
\end{align}
using the $p$-wave propagator for the $\chi^\eta$ field 
\begin{align}
iD_\chi^\eta (p_0,\bm{p})  =& \frac{i   }
{\Delta^{(\eta)} - \frac{1}{2\mu}\zeta^2 +\frac{2[h^{(\eta)}]^2}{\mu}
f_1 (p_0,\bm{p})},\\
f_1(p_0,\bm{p})=&
\frac{1}{4\pi}\left(\zeta^3-\frac{3}{2}\zeta^2\lambda
+\frac{\pi}{2} \lambda^3 \right), \nonumber
\end{align}
where $\zeta=\sqrt{-2\mu p_0 +\mu p^2/M -i 0^+}$. 

The EFT couplings in $p$-wave can be related to observables by comparing the EFT amplitude Eq.~(\ref{eq:p-wave}) to the ERE as done for $s$-wave earlier.  For $p$-wave we get
\begin{align}
i\mathcal A_1^{\eta}(p)= i\frac{2\pi}{\mu}\frac{p^2}{p^3\cot\delta_1^\eta-ip^3}\approx
 i\frac{2\pi}{\mu}\frac{p^2}{-1/a_1^{(\eta)}+r_1^{(\eta)} p^2/2-ip^3} ,
\end{align}
and
\begin{align}\label{eq:a1r1}
-\frac{2\pi\mu\Delta^{(\eta)}}{[h^{(\eta)}]^2}-\frac{\pi}{2}\lambda^3=&
-1/a_1^{(\eta)}, \\
-\frac{3}{2}\lambda-\frac{\pi}{[h^{(\eta)}]^2}=&\frac{1}{2} r_1^{(\eta)}. \nonumber
\end{align}

The ERE parameters $a_1^{(1)}, r_1^{(1)}$  and $a_1^{(2)}, r_1^{(2)}$  can in principle be used to determine the EFT couplings $\Delta^{(1)}, h^{(1)}$ and $\Delta^{(2)}, h^{(2)}$  in the $^2P_{1/2}$  and the $^2P_{3/2}$ channels, respectively.  However, due to lack of sufficient elastic $n+^{14}$C scattering data the ERE parameters in $p$-wave are not known.  In the EFT it is not clear \emph{a priori}  how the couplings should be estimated.  In the natural case where all couplings scale with the short-distance scale $\Lambda$, initial $p$-wave interaction would be perturbative.  In the presence of 
shallow bound, virtual or resonance states in $p$-wave, the EFT couplings are fine tuned to scale with powers of the long-distance scale $Q$. Then the  $p$-wave operators in 
Eq.~(\ref{eq:Lp}) need to be treated non-perturbatively~\cite{Bertulani:2002sz,Bedaque:2003wa}.  Even in the case where $p$-wave interaction is perturbative,  treating it non-perturbatively  does not introduce uncontrolled error in the EFT calculation. Thus resuming the $p$-wave interaction with the interactions in Eq.~(\ref{eq:Lp}) to all order we get a result valid in the natural and un-natural case. 

Out of the four unknown $p$-wave couplings, we can determine two of the couplings from the known resonance $\frac{1}{2}^{-}$ state of $^{15}$C, with a resonance energy $E_r\approx 1.885$ MeV and width $\Gamma_r\approx 40$ keV in the c.m. frame. This resonance state is in the $^2P_{1/2}$ channel in the EFT. To describe the resonance one needs to treat the $p$-wave interaction non-perturbatively. Analysing the elastic scattering amplitude near the resonance, we get~\cite{Fernando:2011ts}
\begin{align}\label{eq:ResonanceParameters}
a_1^{(1)}=-\frac{\mu\Gamma_r}{p_r^5},\ \ \mathrm{and}\ \ r_1^{(1)}=
-\frac{2p_r^3}{\mu\Gamma_r}. 
\end{align}
This determines the couplings $\Delta^{(1)}$, $h^{(1)}$  from the resonance parameters. The $a_1^{(1)}, r_1^{(1)}$ obtained from the $\frac{1}{2}^-$ resonance state when used in the capture cross section 
Eq.~(\ref{eq:E1capture}) gives negligible contribution to \nC~ away from the resonance. Near the resonance it produces a sharp peak as we show later in Fig.~\ref{fig:Resonance}.  We determine  the scaling of the remaining two $p$-wave EFT couplings by analyzing available \nC~  data in the following.

\section{Results}\label{sec_results}

\begin{figure}[thb]
\begin{center}
\includegraphics[width=0.47\textwidth,clip=true]{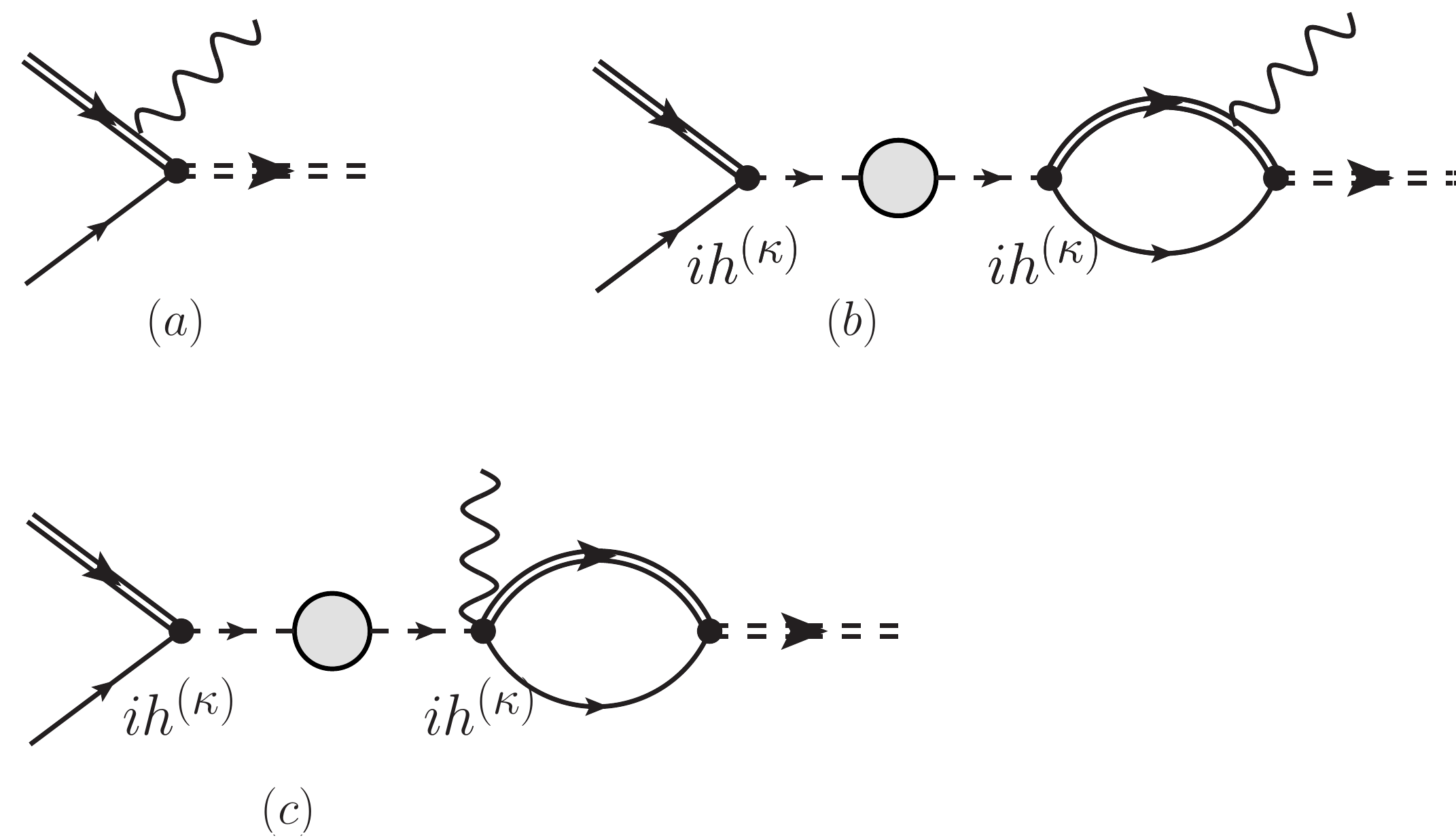} 
\end{center}
\caption{\protect E1 capture.
Double dashed line is used to distinguish the final state $^{15}$C dimer field $\phi$ from the single dashed dressed dimer field $\chi^\eta$ representing initial $p$-wave interaction. Wavy lines represent photons.
 $\kappa=1,2$ corresponds to initial state interaction in the $^2P_{1/2}$ and $^2P_{3/2}$ channels, respectively.}
\label{fig:E1capture}
\end{figure}
The capture reaction \nC~ proceeds through the diagrams in Fig.~\ref{fig:E1capture}. We only concentrate on the E1 transition. The photon couples to the charge of the $^{14}$C core through minimal coupling.  This corresponds to gauging the core momentum $\bm{p}\rightarrow \bm{p}+ Z_c e \bm{A}$, where $Z_c=6$.  The contribution from the first diagram Fig.~\ref{fig:E1capture} $(a)$ can be projected onto capture contribution from initial $^2P_{1/2}$ and $^2P_{3/2}$ channels using the projectors from 
Eq.~(\ref{eq:PP}).  Including the contribution from the diagrams $(b)$ and $(c)$ that involve  the initial state $p$-wave interactions from Eq.~(\ref{eq:Lp}), the amplitude square can be written as
\begin{align}
|\mathcal M^{^2P_{1/2}} |^2= &\left|\frac{12 e h_0\sqrt{Z_\phi}}{M_c}\right|^2 \frac{32 M_n  M_c M p^2}{9}\left| g^{2P_{1/2}}(p)
\right|^2\,\\
g^{^2P_{1/2}}(p)=&\frac{\mu}{p^2+\gamma^2}+\frac{6\pi\mu}{-1/a_1^{(1)}+ r_1^{(1)} p^2/2-ip^3} \[\frac{\gamma}{4\pi}+\frac{ip^3-\gamma^3}{6\pi(p^2+\gamma^2)} \], \nonumber
\end{align}
in the $^2P_{1/2}$ channel. The first term, without the initial state $p$-wave interaction, in $g^{^2P_{1/2}}$ is from diagram  Fig.~\ref{fig:E1capture} $(a)$. In the $^2P_{3/2}$ channel we get a similar expression  
\begin{align}
|\mathcal M^{^2P_{3/2}} |^2= &\left|\frac{12 e h_0\sqrt{Z_\phi}}{M_c}\right|^2 \frac{16 M_n M_cM p^2}{9} \left|  g^{2P_{3/2}}(p)\right|^2 (5-3\cos^2\theta ),  \\
g^{^2P_{3/2}}(p)=&\frac{\mu}{p^2+\gamma^2}+\frac{6\pi\mu}{-1/a_1^{(2)}+ r_1^{(2)} p^2/2-ip^3} \[\frac{\gamma}{4\pi}
+\frac{i p^3-\gamma^3}{6\pi(p^2+\gamma^2)}\]. \nonumber
\end{align}
We used c.m. kinematics: $\bm{p}$ the carbon-14 core momentum, $\bm{k}$ the photon momentum and $\hat{\bm{k}}\cdot\hat{\bm{p}}=\cos\theta$.  
There is a contribution from the interference between the two $p$-wave channels that vanish  when 
we average over the angle $\theta$ to calculate the total unpolarized cross section. 
We made the leading order approximation 
$|\bm{k}|=k_0 \approx (p^2+\gamma^2 )/(2\mu)$.  The wave function renormalization factor $Z_\phi$ is related to the residue at the pole of the propagator of the $\phi$ particle that represents the $^{15}$C ground state. It is calculated from the dressed $\phi$ propagator as
\begin{align}
Z_\phi^{-1} =\frac{\partial}{\partial p_0}[D_\phi(p_0,\bm{p})]^{-1}\Big|_{p_0=p^2/(2M)-B} = 1+\frac{\mu^2 h_0^2}{2\pi\gamma}
=-\frac{1-\rho\gamma}{\rho\gamma}, 
\end{align}
where $B=\gamma^2/(2\mu)\approx 1.218$ MeV is the ground state binding energy. 

The spin averaged differential cross section in c.m. frame is written as
\begin{align}
\frac{d\sigma}{d\cos\theta}=\frac{1}{32\pi s}\frac{|\bm{k}|}{|\bm{p}|}\frac{|\mathcal M |^2}{2}. 
\end{align}
At LO we can write the Mandelstam variable $s\approx (M_n+M_c)^2=M^2$. We write the total 
cross section  as
\begin{align}\label{eq:E1capture}
\sigma (p)=\frac{1}{2}\frac{64\pi\alpha}{M_c^2 \mu^2} \frac{p\gamma(p^2+\gamma^2)}{1-\rho\gamma}
\[ 2 | g^{^2P_{1/2}} (p)|^2 +4 | g^{^2P_{3/2}}(p) |^2\],
\end{align}
where the electron charge is defined as $\alpha= e^2/(4\pi)=1/137$.

\begin{figure}[thb]
\begin{center}
\includegraphics[width=0.46\textwidth,clip=true]{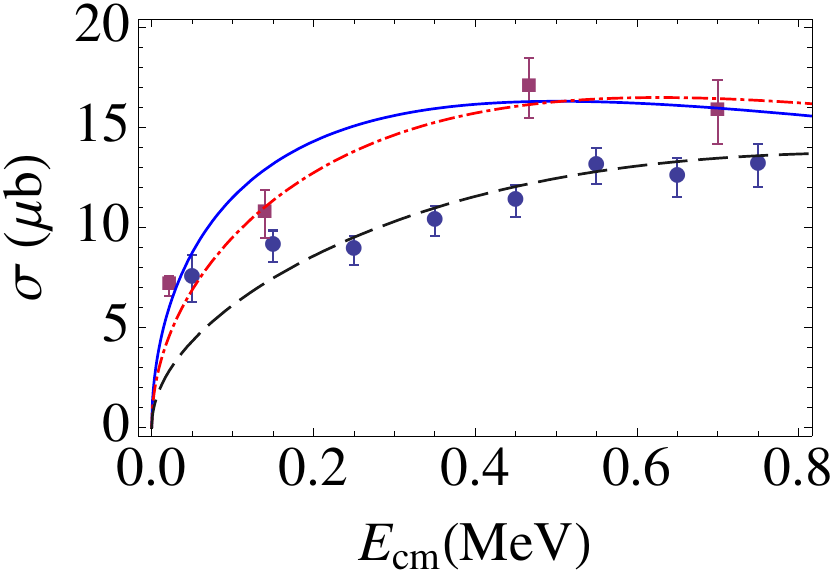}  
\includegraphics[width=0.46\textwidth,clip=true]{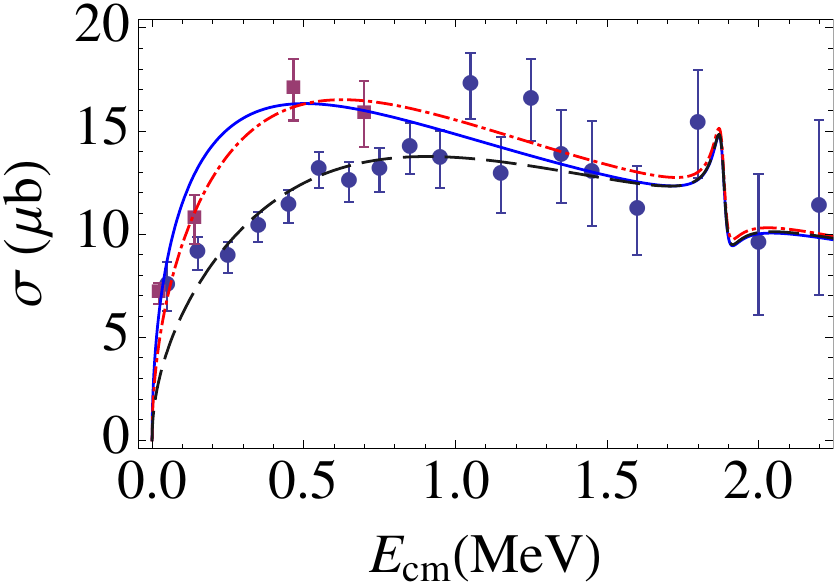} 
\end{center}
\caption{\protect E1 capture cross section $\sigma(E_\mathrm{cm})$ with $a_1^{(2)}=-n_1/(Q^3)$, $r_1^{(2)}=2 n_2  Q$, and  $Q=40$ MeV.  Solid (blue) curve uses  $(n_1,n_2)=$ (2, 1.5); dot-dashed (red) curve uses $(n_1, n_2)=$ (1.5, 1.2); dashed (black) curve uses  $(n_1, n_2)=$ (0.818, 1.12). Square (maroon) direct capture data from Ref.~\cite{Reifarth2008},  
circle (dark blue) Coulomb dissociation data from Ref.~\cite{Nakamura2009}.  }
\label{fig:E1result}
\end{figure}

The cross section in Eq.~(\ref{eq:E1capture}) depends on three unknown EFT couplings  that can be expressed in terms of three ERE parameters: the $s$-wave effective range $\rho$, the $^2P_{3/2}$ channel  scattering volume $a_1^{(2)}$ and the $^2P_{3/2}$ channel ``effective range" $r_1^{(2)}$. Written in this form, the contributions from Figs.~\ref{fig:E1capture} $(a)$, $(b)$ and $(c)$ is model-independent as the ERE parameters are not model specific definitions but universal that are in principle directly related to the $n+^{14}$C elastic scattering phase shifts.   The total $^2P_{1/2}$ and $^2P_{3/2}$ contribution from the tree level diagram Fig.~\ref{fig:E1capture} $(a)$ without the effective range correction $\rho$ is around $5\mu$b.  This is comparable to the data~\cite{Reifarth2008,Nakamura2009} in Fig.~\ref{fig:E1result} but also indicates that effective range $\rho$ correction and/or initial state $p$-wave interaction is important at LO to explain the data.  
 In the natural case $a_1^{(2)}\sim 1/\Lambda^3, r_1^{(2)}\sim \Lambda$, and initial state $p$-wave interaction in Fig.~\ref{fig:E1capture} $(b)$ and $(c)$ is suppressed compared to the diagram $(a)$ by factors of $Q^3/\Lambda^3$.  Two typical unnatural cases in $p$-wave   were considered in Refs.~\cite{Bertulani:2002sz} and \cite{,Bedaque:2003wa}. In the former $a_1^{(2)}\sim 1/Q^3, r_1^{(2)}\sim Q$ and the $p$-wave interaction in all the three diagrams are of the same order. In the latter $a_1^{(2)}\sim 1/(Q^2\Lambda), r_1^{(2)}\sim \Lambda$ and the $p$-wave interaction in diagram $(b)$ and $(c)$ is $Q/\Lambda$ suppressed compare to diagram $(a)$. 
We construct a systematic EFT by considering $\rho\sim 1/\Lambda$ and $a_1^{(2)}\sim 1/Q^3$, $r_1^{(2)}\sim Q$.   Then the  $s$-wave effective range $\rho$  correction is a NLO effect, and the $^2P_{3/2}$ interactions are LO.  We present only the LO result where the effective range $\rho$ contribution is neglected.

\begin{figure}[thb]
\begin{center}
\includegraphics[width=0.46\textwidth,clip=true]{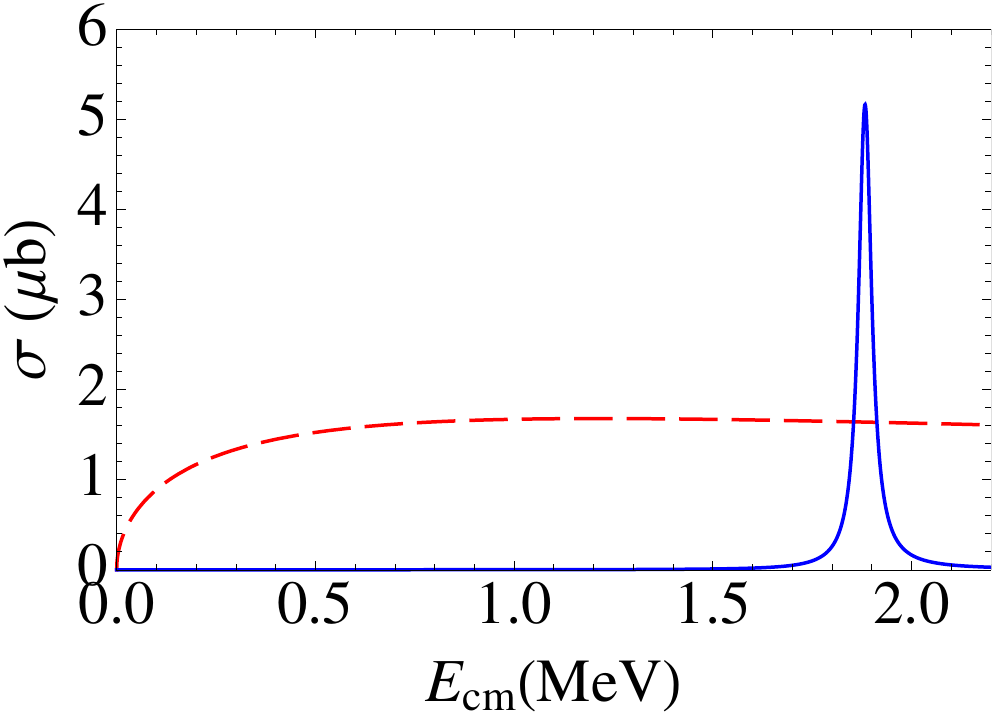}  
\end{center}
\caption{\protect Resonant and non-resonant contribution to E1 capture cross section $\sigma(E_\mathrm{cm})$ in the $^2P_{1/2}$ channel.  Solid (blue) curve is the resonant contribution, and  dashed (red) curve is the non-resonant contribution.  }
\label{fig:Resonance}
\end{figure}
In the $^2P_{1/2}$ channel, the LO cross section is determined by the $^{15}$C ground state binding momentum $\gamma$,  and the $\frac{1}{2}^-$ state resonance energy $E_r$  and width $\Gamma_r$.
In Fig.~\ref{fig:Resonance}, we compare the contribution from Fig.~\ref{fig:E1capture} $(a)$ to that from Fig.~\ref{fig:E1capture} $(b)$, $(c)$. The dashed curve shows the non-resonant contribution in the $^2P_{1/2}$ channel and the solid curve  shows the $\frac{1}{2}^{-}$ resonant contribution (in the same $^2P_{1/2}$ channel). As expected the resonant contribution is large near the resonance energy $E_r\approx 1.885$ MeV, and comparatively negligible elsewhere.  More importantly we notice that the non-resonant contribution is  non-negligible throughout the energy region. This implies that the interference between the resonant and non-resonant contribution in the total cross section is significant as we see later.

In the $^2P_{3/2}$ channel the undetermined ERE parameters are $a_1^{(2)}, r_1^{(2)}$ at LO.
 In Fig.~\ref{fig:E1result} we plot the total cross section  parametrized by 
 $a_1^{(2)}=-n_1/(Q^3)$, $r_1^{(2)}= 2 n_2 Q$   for some reasonable values of $n_1$ and $n_2$ of $\mathcal O(1)$.  
 We pick $Q=40$ MeV. 
  For example, $(n_1,n_2)=$ (2, 1.5) and $(n_1,n_2)=$ (1.5, 1.2)  reproduces direct capture data from Ref.~\cite{Reifarth2008}. 
 We also show Coulomb dissociation data from Ref.~\cite{Nakamura2009}.  
 A $\chi$-square fit to the Coulomb dissociation  data with $Q=40$ MeV gives $(n_1,n_2)=$ (0.818,  1.12). The resonance contribution near $E_\mathrm{cm}\approx 1.89$ MeV differs from a simple Breit-Wigner form. This is a result of the significant interference between the non-resonant and resonant contribution in the $^2P_{1/2}$ channel alluded to earlier in discussing Fig.~\ref{fig:Resonance}.

\begin{figure}[thb]
\begin{center}
\includegraphics[width=0.46\textwidth,clip=true]{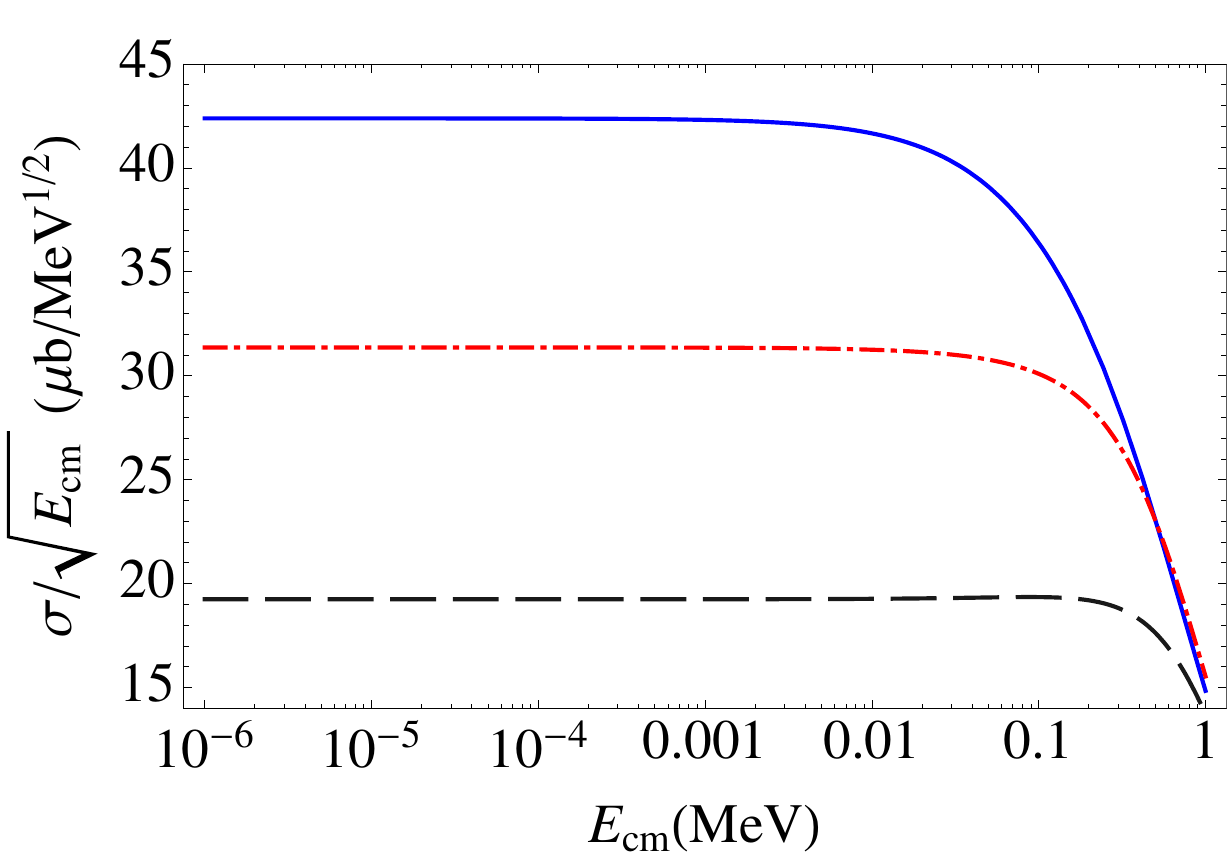}  
\end{center}
\caption{\protect E1 capture $S$-factor $S_n= \sigma / \sqrt{E_\mathrm{cm}}$.  We use the same set of parameters (including $\rho=0$) and legends as 
in Fig.~\ref{fig:E1result}.  
  }
\label{fig:Sfactor}
\end{figure}
Traditionally the cross section $\sigma$ in Eq.~\ref{eq:E1capture} is presented in terms of the $S$-factor $S_n=
\sigma /\sqrt{E_\mathrm{cm}} $ for use in astrophysical calculation at low-energy~\cite{Fowler1967}.  
As the capture proceeds through $p$-wave initial states to $s$-wave final state, the $S$-factor is a constant at low-energy~\cite{Fowler1967,Wiescher1990}. 
In Fig.~\ref{fig:Sfactor} we plot the $S$-factor $S_n= \sigma/ \sqrt{E_\mathrm{cm}} $
using the cross section $\sigma$ from Eq.~(\ref{eq:E1capture}).  We use 
 the same values of parameters (including $\rho=0$)  used in Fig.~\ref{fig:E1result}.  The three set of values for $S_n$ at low-energy are consistent within the $30\%$ accuracy expected of the LO result. 
We note that the larger values of $S_n$ (solid curve) are close to the values obtained in the microscopic calculation in Ref.~\cite{Descouvemont2000}, and the intermediate values of $S_n$ (dot-dashed curve) are close to the values obtained in the potential model calculation in Ref.~\cite{Wiescher1990}.   The $S$-factor is a constant at low-energy and 
expanding  it  to the lowest order in energy we get
\begin{align}\label{eq:Sfactor}
S_n=\frac{16\pi\alpha\sqrt{2\mu}}{M_c^2\gamma(1-\rho\gamma)}
\[12-4(a_1^{(1)}+2 a_1^{(2)})\gamma^3+([a_1^{(1)}]^2+2 [a_1^{(2)}]^2)\gamma^6\]
 +\mathcal O(E_\mathrm{cm}).
\end{align}
The contribution from $p$-wave interaction in the $^2P_{1/2}$ channel through $a_1^{(1)}$ is negligible at low-energy, Fig.~\ref{fig:Resonance}. The result in Eq.~(\ref{eq:Sfactor}) is accurate to NLO at low-energy where contributions from $p$-wave ERE parameters such as $r_1^{(1)}$, $r_1^{(2)}$ are suppressed. 
The NLO correction to $S_n$ at low energy is through the effective range $\rho$ contribution as seen in Eq.~(\ref{eq:Sfactor}).

\begin{figure}[thb]
\begin{center}
\includegraphics[width=0.48\textwidth,clip=true]{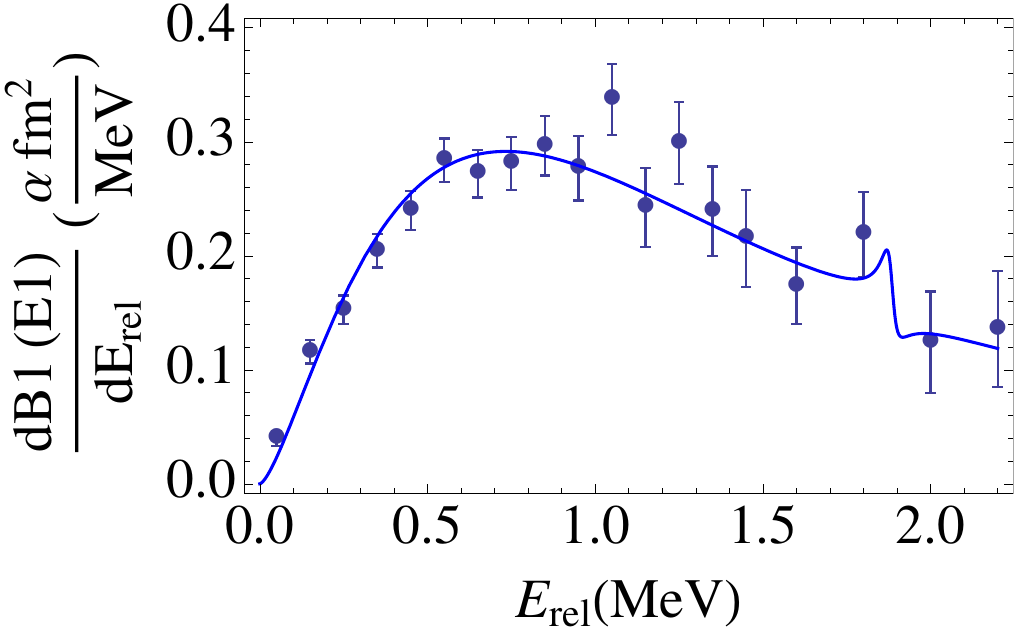}  
\end{center}
\caption{\protect $B(E1)$ strength.  Solid (blue) curve  uses 
$a_1^{(2)}= - 0.818/Q^3$, 
$r_1^{(2)}=1.12\times 2 Q$ with $Q=40$ MeV.  Circle (dark blue) data from  Ref.~\cite{Nakamura2009}.
}
\label{fig:BeE1}
\end{figure}

In Fig.~\ref{fig:BeE1}, we look at the E1 reduced transition probability strength~\cite{Baur:1986pd,Bertulani:2009zk} 
\begin{align}
\frac{d B(E1)}{dE_\mathrm{rel}} = \frac{9}{16\pi^3}\frac{\mu E_\mathrm{cm}}{E_\gamma^3}\sigma(E_\mathrm{cm}),
\label{eq:BeE1}
\end{align}
and compare with available data~\cite{Nakamura2009}.  We ignored any recoil and equated $E_\gamma= E_\mathrm{rel}+B$.
We used $(n_1=0.818, n_2=1.12)$ with $Q=40$ MeV. The agreement  with data is not surprising  since the capture cross section in Fig.~\ref{fig:E1result} was extracted using Eq.~(\ref{eq:BeE1}).  This assumed negligible nuclear contribution from the Pb target at the forward angles (large impact parameter) in Ref.~\cite{Nakamura2009}. 

\section{Conclusions}\label{sec_conclusions}

In this work we consider the radiative capture cross section for \nC~ in halo EFT.   The dominant contribution from E1 transition between initial $p$-wave continuum state and final $s$-wave ground state of $^{15}$C is calculated. The EFT is constructed in the single-particle approximation taking advantage of the low neutron separation energy in  $^{15}$C nuclei. A consistent power counting is developed where the leading contribution involve initial state $p$-wave interactions. 
Both the resonant and non-resonant interaction is considered.

The EFT result is written in a model-independent form using the ERE parameters.  In particular, the result depends on the $^{15}$C ground state binding momentum $\gamma$, and 
 on the scattering parameters $a_1^{(1)}$, $r_1^{(1)}$ and $a_1^{(2)}$, $r_1^{(2)}$ that encapsulate the interactions in the initial $^2P_{1/2}$ and $^2P_{3/2}$ channels, respectively.  
The $^2P_{1/2}$ parameters are constrained using the  resonance energy and width of the $\frac{1}{2}^{-}$ resonance state of $^{15}$C. The scattering parameters in the $^2P_{3/2}$ channel are estimated from direct capture and Coulomb dissociation data.

The EFT calculation is shown to be able to describe the energy dependence of the capture cross section at the order of the calculation. 
The EFT couplings constrained from direct capture reaction and Coulomb dissociation have values consistent with the EFT power counting. The values of the $p$-wave couplings constrained from the direct capture and Coulomb dissociation data are also consistent with each other within the expected EFT error $\mathcal O(Q/\Lambda)\sim 30\%$ on the coupling.   
The contribution from the resonance in the $^2P_{1/2}$ channel differs from a simple Breit-Wigner form due to significant interference with the non-resonant contribution in this channel.   It would be interesting to see if this can be confirmed experimentally with more accurate measurements near the resonance energy.   
Future work should address contributions from the excited $\frac{5}{2}^+$ state of $^{15}$C to the direct capture 
reaction \nC. Higher order contributions from  two-body currents should be explored as well. 

\begin{acknowledgments}
The authors acknowledge helpful discussions with C. Bertulani, H.-W. Hammer, R. Higa and D. R. Phillips.  We thank T. Nakamura for providing the data on Coulomb dissociation. 
This work is partially supported by the U.S. NSF Grant No. PHY-0969378 
and HPC$^2$ Center for
Computational Sciences at Mississippi State University. 
\end{acknowledgments}


\begin{thebibliography}{32}%
\makeatletter
\providecommand \@ifxundefined [1]{%
 \@ifx{#1\undefined}
}%
\providecommand \@ifnum [1]{%
 \ifnum #1\expandafter \@firstoftwo
 \else \expandafter \@secondoftwo
 \fi
}%
\providecommand \@ifx [1]{%
 \ifx #1\expandafter \@firstoftwo
 \else \expandafter \@secondoftwo
 \fi
}%
\providecommand \natexlab [1]{#1}%
\providecommand \enquote  [1]{``#1''}%
\providecommand \bibnamefont  [1]{#1}%
\providecommand \bibfnamefont [1]{#1}%
\providecommand \citenamefont [1]{#1}%
\providecommand \href@noop [0]{\@secondoftwo}%
\providecommand \href [0]{\begingroup \@sanitize@url \@href}%
\providecommand \@href[1]{\@@startlink{#1}\@@href}%
\providecommand \@@href[1]{\endgroup#1\@@endlink}%
\providecommand \@sanitize@url [0]{\catcode `\\12\catcode `\$12\catcode
  `\&12\catcode `\#12\catcode `\^12\catcode `\_12\catcode `\%12\relax}%
\providecommand \@@startlink[1]{}%
\providecommand \@@endlink[0]{}%
\providecommand \url  [0]{\begingroup\@sanitize@url \@url }%
\providecommand \@url [1]{\endgroup\@href {#1}{\urlprefix }}%
\providecommand \urlprefix  [0]{URL }%
\providecommand \Eprint [0]{\href }%
\@ifxundefined \urlstyle {%
  \providecommand \doi  [0]{\begingroup \@sanitize@url \@doi}%
  \providecommand \@doi [1]{\endgroup \@@startlink {\doibase
  #1}doi:\discretionary {}{}{}#1\@@endlink }%
}{%
  \providecommand \doi  [0]{doi:\discretionary{}{}{}\begingroup
  \urlstyle{rm}\Url }%
}%
\providecommand \doibase [0]{http://dx.doi.org/}%
\providecommand \Doi [0]{\begingroup \@sanitize@url \@Doi }%
\providecommand \@Doi  [1]{\endgroup\@@startlink{\doibase#1}\@@Doi}%
\providecommand \@@Doi [1]{#1\@@endlink}%
\providecommand \selectlanguage [0]{\@gobble}%
\providecommand \bibinfo  [0]{\@secondoftwo}%
\providecommand \bibfield  [0]{\@secondoftwo}%
\providecommand \translation [1]{[#1]}%
\providecommand \BibitemOpen [0]{}%
\providecommand \bibitemStop [0]{}%
\providecommand \bibitemNoStop [0]{.\EOS\space}%
\providecommand \EOS [0]{\spacefactor3000\relax}%
\providecommand \BibitemShut  [1]{\csname bibitem#1\endcsname}%
\bibitem [{\citenamefont {Wiescher}\ \emph {et~al.}(1999)\citenamefont
  {Wiescher}, \citenamefont {Gorres},\ and\ \citenamefont
  {Schatz}}]{Wiescher1999}%
  \BibitemOpen
  \bibfield  {author} {\bibinfo {author} {\bibfnamefont {M.}~\bibnamefont
  {Wiescher}}, \bibinfo {author} {\bibfnamefont {J.}~\bibnamefont {Gorres}}, \
  and\ \bibinfo {author} {\bibfnamefont {H.}~\bibnamefont {Schatz}},\
  }\href@noop {} {\bibfield  {journal} {\bibinfo  {journal} {J. Phys. G: Nucl.
  Part. Phys.},\ }\textbf {\bibinfo {volume} {25}},\ \bibinfo {pages} {R133}
  (\bibinfo {year} {1999})}\BibitemShut {NoStop}%
\bibitem [{\citenamefont {Wiescher}\ \emph {et~al.}(1990)\citenamefont
  {Wiescher}, \citenamefont {Gorres},\ and\ \citenamefont
  {Thielemann}}]{Wiescher1990}%
  \BibitemOpen
  \bibfield  {author} {\bibinfo {author} {\bibfnamefont {M.}~\bibnamefont
  {Wiescher}}, \bibinfo {author} {\bibfnamefont {J.}~\bibnamefont {Gorres}}, \
  and\ \bibinfo {author} {\bibfnamefont {F.-K.}\ \bibnamefont {Thielemann}},\
  }\href@noop {} {\bibfield  {journal} {\bibinfo  {journal} {Astrophys J.},\
  }\textbf {\bibinfo {volume} {363}},\ \bibinfo {pages} {340} (\bibinfo {year}
  {1990})}\BibitemShut {NoStop}%
\bibitem [{\citenamefont {Kajino}\ \emph {et~al.}()\citenamefont {Kajino},
  \citenamefont {Mathews},\ and\ \citenamefont {Fuller}}]{Kajino1990}%
  \BibitemOpen
  \bibfield  {author} {\bibinfo {author} {\bibfnamefont {T.}~\bibnamefont
  {Kajino}}, \bibinfo {author} {\bibfnamefont {G.~J.}\ \bibnamefont {Mathews}},
  \ and\ \bibinfo {author} {\bibfnamefont {G.~M.}\ \bibnamefont {Fuller}},\
  }\href@noop {} {}\bibinfo {note} {Astrophys. J. 364, 7 (1990)}\BibitemShut
  {NoStop}%
\bibitem [{\citenamefont {Beer}\ \emph {et~al.}(1992)\citenamefont {Beer} \emph
  {et~al.}}]{Beer1992}%
  \BibitemOpen
  \bibfield  {author} {\bibinfo {author} {\bibfnamefont {H.}~\bibnamefont
  {Beer}} \emph {et~al.},\ }\href@noop {} {\bibfield  {journal} {\bibinfo
  {journal} {Astrophys. J.},\ }\textbf {\bibinfo {volume} {387}},\ \bibinfo
  {pages} {258} (\bibinfo {year} {1992})}\BibitemShut {NoStop}%
\bibitem [{\citenamefont {Reifarth}\ \emph {et~al.}(2005)\citenamefont
  {Reifarth} \emph {et~al.}}]{Reifarth2005}%
  \BibitemOpen
  \bibfield  {author} {\bibinfo {author} {\bibfnamefont {R.}~\bibnamefont
  {Reifarth}} \emph {et~al.},\ }\href@noop {} {\bibfield  {journal} {\bibinfo
  {journal} {Nucl. Phys. A},\ }\textbf {\bibinfo {volume} {758}},\ \bibinfo
  {pages} {787c} (\bibinfo {year} {2005})}\BibitemShut {NoStop}%
\bibitem [{\citenamefont {Reifarth}\ \emph {et~al.}(2008)\citenamefont
  {Reifarth} \emph {et~al.}}]{Reifarth2008}%
  \BibitemOpen
  \bibfield  {author} {\bibinfo {author} {\bibfnamefont {R.}~\bibnamefont
  {Reifarth}} \emph {et~al.},\ }\Doi {10.1103/PhysRevC.77.015804} {\bibfield
  {journal} {\bibinfo  {journal} {Phys. Rev. C},\ }\textbf {\bibinfo {volume}
  {77}},\ \bibinfo {pages} {015804} (\bibinfo {year} {2008})}\BibitemShut
  {NoStop}%
\bibitem [{\citenamefont {Horv\'ath}\ \emph {et~al.}(2002)\citenamefont
  {Horv\'ath} \emph {et~al.}}]{Horvath2002}%
  \BibitemOpen
  \bibfield  {author} {\bibinfo {author} {\bibfnamefont {A.}~\bibnamefont
  {Horv\'ath}} \emph {et~al.},\ }\href@noop {} {\bibfield  {journal} {\bibinfo
  {journal} {Astrophys. J.},\ }\textbf {\bibinfo {volume} {570}},\ \bibinfo
  {pages} {926} (\bibinfo {year} {2002})}\BibitemShut {NoStop}%
\bibitem [{\citenamefont {Datta~Pramanik}\ \emph {et~al.}(2003)\citenamefont
  {Datta~Pramanik} \emph {et~al.}}]{Pramanik2003}%
  \BibitemOpen
  \bibfield  {author} {\bibinfo {author} {\bibfnamefont {U.}~\bibnamefont
  {Datta~Pramanik}} \emph {et~al.},\ }\href@noop {} {\bibfield  {journal}
  {\bibinfo  {journal} {Phys. Lett. B},\ }\textbf {\bibinfo {volume} {551}},\
  \bibinfo {pages} {63} (\bibinfo {year} {2003})}\BibitemShut {NoStop}%
\bibitem [{\citenamefont {Nakamura}\ \emph {et~al.}(2003)\citenamefont
  {Nakamura} \emph {et~al.}}]{Nakamura2003}%
  \BibitemOpen
  \bibfield  {author} {\bibinfo {author} {\bibfnamefont {T.}~\bibnamefont
  {Nakamura}} \emph {et~al.},\ }\href@noop {} {\bibfield  {journal} {\bibinfo
  {journal} {Nucl. Phys. A},\ }\textbf {\bibinfo {volume} {722}},\ \bibinfo
  {pages} {301} (\bibinfo {year} {2003})}\BibitemShut {NoStop}%
\bibitem [{\citenamefont {Nakamura}\ \emph {et~al.}(2009)\citenamefont
  {Nakamura} \emph {et~al.}}]{Nakamura2009}%
  \BibitemOpen
  \bibfield  {author} {\bibinfo {author} {\bibfnamefont {T.}~\bibnamefont
  {Nakamura}} \emph {et~al.},\ }\Doi {10.1103/PhysRevC.79.035805} {\bibfield
  {journal} {\bibinfo  {journal} {Phys. Rev. C},\ }\textbf {\bibinfo {volume}
  {79}},\ \bibinfo {pages} {035805} (\bibinfo {year} {2009})}\BibitemShut
  {NoStop}%
\bibitem [{\citenamefont {Baur}\ \emph {et~al.}(1986)\citenamefont {Baur},
  \citenamefont {Bertulani},\ and\ \citenamefont {Rebel}}]{Baur:1986pd}%
  \BibitemOpen
  \bibfield  {author} {\bibinfo {author} {\bibfnamefont {G.}~\bibnamefont
  {Baur}}, \bibinfo {author} {\bibfnamefont {C.}~\bibnamefont {Bertulani}}, \
  and\ \bibinfo {author} {\bibfnamefont {H.}~\bibnamefont {Rebel}},\ }\Doi
  {10.1016/0375-9474(86)90290-3} {\bibfield  {journal} {\bibinfo  {journal}
  {Nucl.Phys.},\ }\textbf {\bibinfo {volume} {A458}},\ \bibinfo {pages} {188}
  (\bibinfo {year} {1986})}\BibitemShut {NoStop}%
\bibitem [{\citenamefont {Summers}\ and\ \citenamefont
  {Nunes}(2008)}]{PhysRevC.78.011601}%
  \BibitemOpen
  \bibfield  {author} {\bibinfo {author} {\bibfnamefont {N.~C.}\ \bibnamefont
  {Summers}}\ and\ \bibinfo {author} {\bibfnamefont {F.~M.}\ \bibnamefont
  {Nunes}},\ }\Doi {10.1103/PhysRevC.78.011601} {\bibfield  {journal} {\bibinfo
   {journal} {Phys. Rev. C},\ }\textbf {\bibinfo {volume} {78}},\ \bibinfo
  {pages} {011601} (\bibinfo {year} {2008})}\BibitemShut {NoStop}%
\bibitem [{\citenamefont {Esbensen}(2009)}]{PhysRevC.80.024608}%
  \BibitemOpen
  \bibfield  {author} {\bibinfo {author} {\bibfnamefont {H.}~\bibnamefont
  {Esbensen}},\ }\Doi {10.1103/PhysRevC.80.024608} {\bibfield  {journal}
  {\bibinfo  {journal} {Phys. Rev. C},\ }\textbf {\bibinfo {volume} {80}},\
  \bibinfo {pages} {024608} (\bibinfo {year} {2009})}\BibitemShut {NoStop}%
\bibitem [{\citenamefont {Timofeyuk}\ \emph {et~al.}(2006)\citenamefont
  {Timofeyuk}, \citenamefont {Baye}, \citenamefont {Descouvemont},
  \citenamefont {Kamouni},\ and\ \citenamefont
  {Thompson}}]{PhysRevLett.96.162501}%
  \BibitemOpen
  \bibfield  {author} {\bibinfo {author} {\bibfnamefont {N.~K.}\ \bibnamefont
  {Timofeyuk}}, \bibinfo {author} {\bibfnamefont {D.}~\bibnamefont {Baye}},
  \bibinfo {author} {\bibfnamefont {P.}~\bibnamefont {Descouvemont}}, \bibinfo
  {author} {\bibfnamefont {R.}~\bibnamefont {Kamouni}}, \ and\ \bibinfo
  {author} {\bibfnamefont {I.~J.}\ \bibnamefont {Thompson}},\ }\Doi
  {10.1103/PhysRevLett.96.162501} {\bibfield  {journal} {\bibinfo  {journal}
  {Phys. Rev. Lett.},\ }\textbf {\bibinfo {volume} {96}},\ \bibinfo {pages}
  {162501} (\bibinfo {year} {2006})}\BibitemShut {NoStop}%
\bibitem [{FRI()}]{FRIB}%
  \BibitemOpen
  \href@noop {} {}\bibinfo {note} {{The Facility for Rare Isotope Beams (FRIB)
  at the Michigan State University, http://frib.msu.edu/}}\BibitemShut
  {NoStop}%
\bibitem [{\citenamefont {Bertulani}\ \emph {et~al.}(2002)\citenamefont
  {Bertulani}, \citenamefont {Hammer},\ and\ \citenamefont
  {Van~Kolck}}]{Bertulani:2002sz}%
  \BibitemOpen
  \bibfield  {author} {\bibinfo {author} {\bibfnamefont {C.~A.}\ \bibnamefont
  {Bertulani}}, \bibinfo {author} {\bibfnamefont {H.~W.}\ \bibnamefont
  {Hammer}}, \ and\ \bibinfo {author} {\bibfnamefont {U.}~\bibnamefont
  {Van~Kolck}},\ }\Doi {10.1016/S0375-9474(02)01270-8} {\bibfield  {journal}
  {\bibinfo  {journal} {Nucl. Phys.},\ }\textbf {\bibinfo {volume} {A712}},\
  \bibinfo {pages} {37} (\bibinfo {year} {2002})}\BibitemShut {NoStop}%
\bibitem [{\citenamefont {Bedaque}\ \emph {et~al.}(2003)\citenamefont
  {Bedaque}, \citenamefont {Hammer},\ and\ \citenamefont {van
  Kolck}}]{Bedaque:2003wa}%
  \BibitemOpen
  \bibfield  {author} {\bibinfo {author} {\bibfnamefont {P.~F.}\ \bibnamefont
  {Bedaque}}, \bibinfo {author} {\bibfnamefont {H.~W.}\ \bibnamefont {Hammer}},
  \ and\ \bibinfo {author} {\bibfnamefont {U.}~\bibnamefont {van Kolck}},\
  }\Doi {10.1016/j.physletb.2003.07.049} {\bibfield  {journal} {\bibinfo
  {journal} {Phys. Lett.},\ }\textbf {\bibinfo {volume} {B569}},\ \bibinfo
  {pages} {159} (\bibinfo {year} {2003})}\BibitemShut {NoStop}%
\bibitem [{\citenamefont {Huang}\ \emph {et~al.}(2010)\citenamefont {Huang},
  \citenamefont {Bertulani},\ and\ \citenamefont {Guimaraes}}]{Huang:2008ye}%
  \BibitemOpen
  \bibfield  {author} {\bibinfo {author} {\bibfnamefont {J.~T.}\ \bibnamefont
  {Huang}}, \bibinfo {author} {\bibfnamefont {C.~A.}\ \bibnamefont
  {Bertulani}}, \ and\ \bibinfo {author} {\bibfnamefont {V.}~\bibnamefont
  {Guimaraes}},\ }\href@noop {} {\bibfield  {journal} {\bibinfo  {journal} {At.
  Data Nuc. Data Tables},\ }\textbf {\bibinfo {volume} {96}},\ \bibinfo {pages}
  {824} (\bibinfo {year} {2010})}\BibitemShut {NoStop}%
\bibitem [{\citenamefont {Descouvemont}(2000)}]{Descouvemont2000}%
  \BibitemOpen
  \bibfield  {author} {\bibinfo {author} {\bibfnamefont {P.}~\bibnamefont
  {Descouvemont}},\ }\href@noop {} {\bibfield  {journal} {\bibinfo  {journal}
  {Nucl. Phys. A},\ }\textbf {\bibinfo {volume} {675}},\ \bibinfo {pages} {559}
  (\bibinfo {year} {2000})}\BibitemShut {NoStop}%
\bibitem [{\citenamefont {Higa}\ \emph {et~al.}(2008)\citenamefont {Higa},
  \citenamefont {Hammer},\ and\ \citenamefont {van Kolck}}]{Higa:2008dn}%
  \BibitemOpen
  \bibfield  {author} {\bibinfo {author} {\bibfnamefont {R.}~\bibnamefont
  {Higa}}, \bibinfo {author} {\bibfnamefont {H.-W.}\ \bibnamefont {Hammer}}, \
  and\ \bibinfo {author} {\bibfnamefont {U.}~\bibnamefont {van Kolck}},\ }\Doi
  {10.1016/j.nuclphysa.2008.06.003} {\bibfield  {journal} {\bibinfo  {journal}
  {Nucl. Phys. A},\ }\textbf {\bibinfo {volume} {809}},\ \bibinfo {pages} {171}
  (\bibinfo {year} {2008})},\ \Eprint {http://arxiv.org/abs/0802.3426}
  {arXiv:0802.3426 [nucl-th]} \BibitemShut {NoStop}%
\bibitem [{\citenamefont {Gelman}(2009)}]{Gelman:2009be}%
  \BibitemOpen
  \bibfield  {author} {\bibinfo {author} {\bibfnamefont {B.~A.}\ \bibnamefont
  {Gelman}},\ }\Doi {10.1103/PhysRevC.80.034005} {\bibfield  {journal}
  {\bibinfo  {journal} {Phys. Rev. C},\ }\textbf {\bibinfo {volume} {80}},\
  \bibinfo {pages} {034005} (\bibinfo {year} {2009})},\ \Eprint
  {http://arxiv.org/abs/0906.5502} {arXiv:0906.5502 [nucl-th]} \BibitemShut
  {NoStop}%
\bibitem [{\citenamefont {Canham}\ and\ \citenamefont
  {Hammer}(2008)}]{Canham:2008jd}%
  \BibitemOpen
  \bibfield  {author} {\bibinfo {author} {\bibfnamefont {D.~L.}\ \bibnamefont
  {Canham}}\ and\ \bibinfo {author} {\bibfnamefont {H.-W.}\ \bibnamefont
  {Hammer}},\ }\Doi {10.1140/epja/i2008-10632-4} {\bibfield  {journal}
  {\bibinfo  {journal} {Eur. Phys. J. A},\ }\textbf {\bibinfo {volume} {37}},\
  \bibinfo {pages} {367} (\bibinfo {year} {2008})},\ \Eprint
  {http://arxiv.org/abs/0807.3258} {arXiv:0807.3258 [nucl-th]} \BibitemShut
  {NoStop}%
\bibitem [{\citenamefont {Canham}\ and\ \citenamefont
  {Hammer}(2010)}]{Canham:2009xg}%
  \BibitemOpen
  \bibfield  {author} {\bibinfo {author} {\bibfnamefont {D.~L.}\ \bibnamefont
  {Canham}}\ and\ \bibinfo {author} {\bibfnamefont {H.-W.}\ \bibnamefont
  {Hammer}},\ }\Doi {10.1016/j.nuclphysa.2010.02.014} {\bibfield  {journal}
  {\bibinfo  {journal} {Nucl. Phys. A},\ }\textbf {\bibinfo {volume} {836}},\
  \bibinfo {pages} {275} (\bibinfo {year} {2010})},\ \Eprint
  {http://arxiv.org/abs/0911.3238} {arXiv:0911.3238 [nucl-th]} \BibitemShut
  {NoStop}%
\bibitem [{\citenamefont {Fernando}\ \emph {et~al.}(2012)\citenamefont
  {Fernando}, \citenamefont {Higa},\ and\ \citenamefont
  {Rupak}}]{Fernando:2011ts}%
  \BibitemOpen
  \bibfield  {author} {\bibinfo {author} {\bibfnamefont {L.}~\bibnamefont
  {Fernando}}, \bibinfo {author} {\bibfnamefont {R.}~\bibnamefont {Higa}}, \
  and\ \bibinfo {author} {\bibfnamefont {G.}~\bibnamefont {Rupak}},\ }\Doi
  {10.1140/epja/i2012-12024-7} {\bibfield  {journal} {\bibinfo  {journal} {Eur.
  Phys. J A},\ }\textbf {\bibinfo {volume} {48}},\ \bibinfo {pages} {24}
  (\bibinfo {year} {2012})}\BibitemShut {NoStop}%
\bibitem [{\citenamefont {Rupak}\ and\ \citenamefont
  {Higa}(2011)}]{Rupak:2011nk}%
  \BibitemOpen
  \bibfield  {author} {\bibinfo {author} {\bibfnamefont {G.}~\bibnamefont
  {Rupak}}\ and\ \bibinfo {author} {\bibfnamefont {R.}~\bibnamefont {Higa}},\
  }\Doi {10.1103/PhysRevLett.106.222501} {\bibfield  {journal} {\bibinfo
  {journal} {Phys. Rev. Lett.},\ }\textbf {\bibinfo {volume} {106}},\ \bibinfo
  {pages} {222501} (\bibinfo {year} {2011})},\ \Eprint
  {http://arxiv.org/abs/1101.0207} {arXiv:1101.0207 [nucl-th]} \BibitemShut
  {NoStop}%
\bibitem [{\citenamefont {Hammer}\ and\ \citenamefont
  {Phillips}(2011)}]{Hammer:2011ye}%
  \BibitemOpen
  \bibfield  {author} {\bibinfo {author} {\bibfnamefont {H.-W.}\ \bibnamefont
  {Hammer}}\ and\ \bibinfo {author} {\bibfnamefont {D.}~\bibnamefont
  {Phillips}},\ }\Doi {10.1016/j.nuclphysa.2011.06.028} {\bibfield  {journal}
  {\bibinfo  {journal} {Nucl. Phys. A},\ }\textbf {\bibinfo {volume} {865}},\
  \bibinfo {pages} {17} (\bibinfo {year} {2011})},\ \Eprint
  {http://arxiv.org/abs/1103.1087} {arXiv:1103.1087 [nucl-th]} \BibitemShut
  {NoStop}%
\bibitem [{\citenamefont {Phillips}\ and\ \citenamefont
  {Hammer}(2010)}]{Phillips:2010dt}%
  \BibitemOpen
  \bibfield  {author} {\bibinfo {author} {\bibfnamefont {D.~R.}\ \bibnamefont
  {Phillips}}\ and\ \bibinfo {author} {\bibfnamefont {H.~W.}\ \bibnamefont
  {Hammer}},\ }\href@noop {} {\bibfield  {journal} {\bibinfo  {journal} {EPJ
  Web Conf.},\ }\textbf {\bibinfo {volume} {3}},\ \bibinfo {pages} {06002}
  (\bibinfo {year} {2010})}\BibitemShut {NoStop}%
\bibitem [{\citenamefont {Lensky}\ and\ \citenamefont
  {Birse}(2011)}]{Lensky:2011he}%
  \BibitemOpen
  \bibfield  {author} {\bibinfo {author} {\bibfnamefont {V.}~\bibnamefont
  {Lensky}}\ and\ \bibinfo {author} {\bibfnamefont {M.~C.}\ \bibnamefont
  {Birse}},\ }\Doi {10.1140/epja/i2011-11142-0} {\bibfield  {journal} {\bibinfo
   {journal} {Eur. Phys. J. A},\ }\textbf {\bibinfo {volume} {47}},\ \bibinfo
  {pages} {142} (\bibinfo {year} {2011})},\ \Eprint
  {http://arxiv.org/abs/1109.2797} {arXiv:1109.2797 [nucl-th]} \BibitemShut
  {NoStop}%
\bibitem [{\citenamefont {Wang}\ \emph {et~al.}(2009)\citenamefont {Wang},
  \citenamefont {Ciss\'e},\ and\ \citenamefont {Baye}}]{PhysRevC.80.034611}%
  \BibitemOpen
  \bibfield  {author} {\bibinfo {author} {\bibfnamefont {C.}~\bibnamefont
  {Wang}}, \bibinfo {author} {\bibfnamefont {O.~I.}\ \bibnamefont {Ciss\'e}}, \
  and\ \bibinfo {author} {\bibfnamefont {D.}~\bibnamefont {Baye}},\ }\Doi
  {10.1103/PhysRevC.80.034611} {\bibfield  {journal} {\bibinfo  {journal}
  {Phys. Rev. C},\ }\textbf {\bibinfo {volume} {80}},\ \bibinfo {pages}
  {034611} (\bibinfo {year} {2009})}\BibitemShut {NoStop}%
\bibitem [{\citenamefont {Kaplan}\ \emph {et~al.}(1998)\citenamefont {Kaplan},
  \citenamefont {Savage},\ and\ \citenamefont {Wise}}]{Kaplan:1998tg}%
  \BibitemOpen
  \bibfield  {author} {\bibinfo {author} {\bibfnamefont {D.~B.}\ \bibnamefont
  {Kaplan}}, \bibinfo {author} {\bibfnamefont {M.~J.}\ \bibnamefont {Savage}},
  \ and\ \bibinfo {author} {\bibfnamefont {M.~B.}\ \bibnamefont {Wise}},\ }\Doi
  {10.1016/S0370-2693(98)00210-X} {\bibfield  {journal} {\bibinfo  {journal}
  {Phys. Lett.},\ }\textbf {\bibinfo {volume} {B424}},\ \bibinfo {pages} {390}
  (\bibinfo {year} {1998})}\BibitemShut {NoStop}%
\bibitem [{\citenamefont {{Fowler}}\ \emph {et~al.}(1967)\citenamefont
  {{Fowler}}, \citenamefont {{Caughlan}},\ and\ \citenamefont
  {{Zimmerman}}}]{Fowler1967}%
  \BibitemOpen
  \bibfield  {author} {\bibinfo {author} {\bibfnamefont {W.~A.}\ \bibnamefont
  {{Fowler}}}, \bibinfo {author} {\bibfnamefont {G.~R.}\ \bibnamefont
  {{Caughlan}}}, \ and\ \bibinfo {author} {\bibfnamefont {B.~A.}\ \bibnamefont
  {{Zimmerman}}},\ }\Doi {10.1146/annurev.aa.05.090167.002521} {\bibfield
  {journal} {\bibinfo  {journal} {Ann. Rev. Astr. Astrophys.},\ }\textbf
  {\bibinfo {volume} {5}},\ \bibinfo {pages} {525} (\bibinfo {year}
  {1967})}\BibitemShut {NoStop}%
\bibitem [{\citenamefont {Bertulani}(2009)}]{Bertulani:2009zk}%
  \BibitemOpen
  \bibfield  {author} {\bibinfo {author} {\bibfnamefont {C.}~\bibnamefont
  {Bertulani}},\ }\href@noop {} { (\bibinfo {year} {2009})},\ \Eprint
  {http://arxiv.org/abs/0908.4307} {arXiv:0908.4307 [nucl-th]} \BibitemShut
  {NoStop}%
\end{thebibliography}

%
 
\end{document}